\documentclass[showpacs,10pt,twocolumn,prb]{revtex4-1}
\usepackage{amsmath}
\usepackage{amssymb}
\usepackage{graphics}
\usepackage{epsfig}
\usepackage{CJK}

\begin{document}

\begin{CJK*}{GBK}{}

\title{Multiband effects on $\beta $-FeSe single crystals}
\author{Hechang Lei,$^{1}$ D. Graf,$^{2}$ Rongwei Hu,$^{1, \ast }$ Hyejin
Ryu,$^{1,3}$ E. S. Choi,$^{2}$ S. W. Tozer,$^{2}$ and C. Petrovic$^{1,3}$}
\date{\today}

\affiliation{$^{1}$Condensed Matter Physics and Materials Science Department, Brookhaven
National Laboratory, Upton, NY 11973, USA}
\affiliation{$^{2}$NHMFL/Physics,
Florida State University, Tallahassee, Florida 32310, USA}
\affiliation{$^{3}$Department of Physics and Astronomy, State University of
New York at Stony Brook, Stony Brook, NY 11794, USA}

\begin{abstract}
We present the upper critical fields $\mu _{0}H_{c2}(T)$ and Hall effect in
$\beta $-FeSe single crystals. The $\mu _{0}H_{c2}(T)$ increases as the
temperature is lowered for field applied parallel and perpendicular to
(101), the natural growth facet of the crystal. The $\mu _{0}H_{c2}(T)$ for
both field directions and the anisotropy at low temperature increase under
pressure. Hole carriers are dominant at high magnetic fields. However, the
contribution of electron-type carriers is significant at low fields and low
temperature. Our results show that multiband effects dominate $\mu
_{0}H_{c2}(T)$ and electronic transport in the normal state.
\end{abstract}

\maketitle

\end{CJK*}

\section{Introduction}

The discovery of iron-based superconductors has\ generated a great deal of
interests because of rather high transition temperature $T_{c}$ and high
upper critical fields $\mu _{0}H_{c2}$. Crystal structures of iron-based
superconductors can be mainly categorized into several types: FePn-1111 type
(REOFePn, RE = rare earth; Pn = P or As), FePn-122 type (AFe$_{2}$As$_{2}$,
A = alkaline or alkaline-earth metals), FePn111 type (AFeAs), FeCh-11 type
(FeCh, Ch = S, Se, Te), FeCh-122 type (A$_{x}$Fe$_{2-y}$Ch$_{2}$), and other
structures with more complex oxide layers.\cite{Kamihara}$^{-}$\cite{Guo}
Despite structural similarity, i.e., shared FePn or FeCh tetrahedron layers,
iron-based superconductors exhibit diverse physical properties. These
include possible differences in pairing symmetry,\cite{Mazin}$^{-}$\cite%
{Zhang} relation to competing or coexisting order states (spin density wave
vs. superconductivity),\cite{Cruz}$^{-}$\cite{Chen H} and diverse normal
state properties.\cite{Kamihara}$^{,}$\cite{Guo} FeCh-11 type materials are
of special interest because their crystal structure has no blocking layers
in between FeCh layers, yet they have similar calculated Fermi surface
topology when compared to other iron-based superconductors.\cite{Subedi}
Furthermore, they also exhibit some exotic features, such as significant
pressure effect,\cite{Mizuguchi2}$^{,}$\cite{Medvedev} and excess Fe with
local moment according to theoretical calculation.\cite{Zhang LJ}

The $\mu _{0}H_{c2}$ gives some important information on fundamental
superconducting properties: coherence length, anisotropy, details of
underlying electronic structures and dimensionality of superconductivity as
well as insights into the pair-breaking mechanism. Previous studies on FeTe$%
_{1-x}$Se$_{x}$ and FeTe$_{1-x}$S$_{x}$ single crystals indicate that the
spin-paramagnetic effect is the main pair-breaking mechanism.\cite{Kida}$^{-}
$\cite{Lei HC2} However, for FePn-1111 and FePn-122 type superconductors the
two-band effect with high diffusivity ratio between different bands
dominates $\mu _{0}H_{c2}(T)$.\cite{Hunte}$^{-}$\cite{Baily}

On the other hand, magnetic penetration depth study of $\beta $-FeSe
polycrystal indicates that $\beta $-FeSe is a two-band superconductor.\cite%
{Khasanov} Therefore, it is of interest to investigate multiband and spin
paramagnetic effects on the $\mu _{0}H_{c2}$ of $\beta $-FeSe. An extremely
complex binary alloy phase diagram and associated difficulties in single
crystal preparation impeded the growth of pure $\beta $-FeSe single crystals.%
\cite{Hu RW} Hence, systematic studies of anisotropy in $\mu _{0}H_{c2}(T)$
and pair breaking mechanism in high magnetic field are still lacking.

In this work, we report on the upper critical fields of pure $\beta $-FeSe
single crystals in dc high magnetic fields up to 35 T at ambient and high
pressures. The results shows that two-band features dominate the pair
breaking with additional influence of spin paramagnetic effect.

\section{Experiment}

Details of crystal synthesis and characterization are explained elsewhere.%
\cite{Hu RW} The $\mu _{0}H_{c2}$ is determined by measuring the magnetic
field dependence of radio frequency (rf) contactless penetration depth in a
static magnet up to 35 T at the National High Magnetic Field Laboratory
(NHMFL) in Tallahassee, Florida. The rf technique is very sensitive to small
changes in the rf penetration depth (about 1-5 nm) in the mixed state and
thus is an accurate method for determining the $\mu _{0}H_{c2}$ of
superconductors.\cite{Mielke} At certain magnetic field, the probe detects
the transition to the normal state by tracking the shift in resonant
frequency, which is proportional to the change in penetration depth as $%
\Delta \lambda $ $\varpropto $ $\Delta F$. Small single crystals were chosen
and the sample was placed in a circular detection coil. More details can be
found in Refs. 29 and 30. For measurement under pressure, the sample was
placed in a 15 turn coil within the gasket hole of a turnbuckle diamond
anvil cell (DAC) made of beryllium copper and containing diamonds with 1.2
mm culets.\cite{Graf} The pressure was calibrated at $\sim $ 4 K by
comparing the fluorescence of a small chip of ruby within the DAC with an
ambient ruby at the same temperature.\cite{Barnett} The small dimensions of
the DAC allow for angular rotation with respect to the applied magnetic
field so H$\parallel $(101) and H$\perp $(101) orientations can be explored
\textit{in situ}. Using four-probe configuration of Hall measurement, the
Hall resistivity was extracted from the difference of transverse resistance
measured at the positive and negative fields, i.e., $\rho _{xy}(H)=[\rho
(+H)-\rho (-H)]/2$, which can effectively eliminate the longitudinal
resistivity component due to voltage probe misalignment.

\section{Results and Discussion}

\begin{figure}[tbp]
\centerline{\includegraphics[scale=0.52]{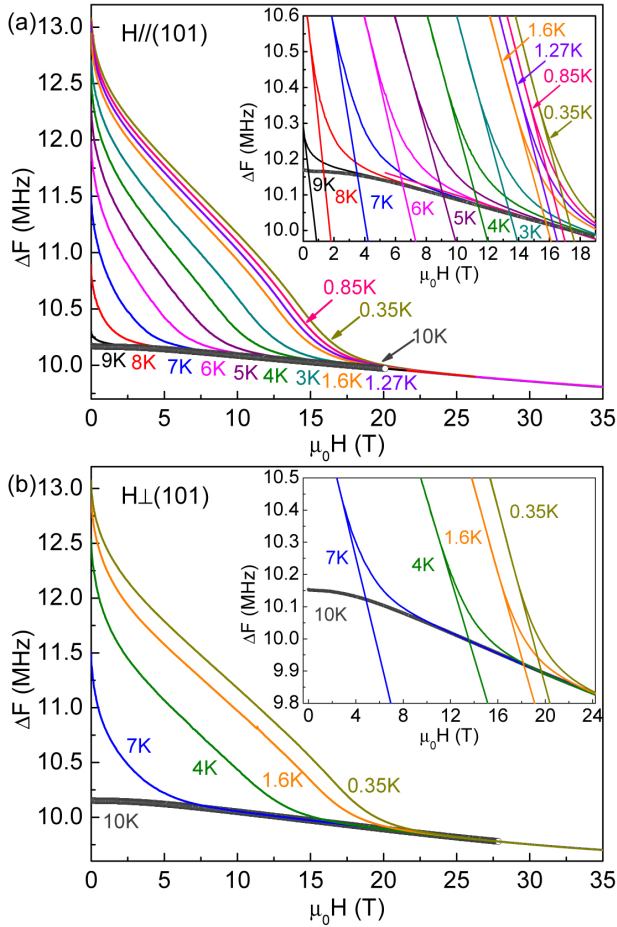}} \vspace*{-0.3cm}
\caption{Field dependence of frequency shift ($\Delta F$) for (a) H$%
\parallel $(101) and (b) H$\perp $(101) at various temperatures. $\Delta F$($%
T$ = 10 K) is set as a normal-state background signal. Inset (a) and (b):
enlarged parts near the points that deviate from background signals. $%
\protect\mu _{0}H_{c2}(T)$ are determined from the interceptions between
extrapolated straight lines below inflexion points and the curves of
normal-state backgrounds for H$\parallel $(101) and H$\perp $(101),
respectively. In the inset (a), the intersection of two slopes in $\Delta F$(%
$T$ = 6 K) curve exhibits another criterion to determine the $\protect\mu %
_{0}H_{c2}(T)$ and the difference between two criterions is taken as the
error bar of $\protect\mu _{0}H_{c2}(T)$.}
\end{figure}

As shown in the main panel of Fig. 1(a) and (b), the rf shift ($\Delta F$)
at 10 K (above $T_{c}$) shows a smooth and almost linear magnetic-field
dependence without any steep changes for both field directions. In the
normal state the rf shift is sensitive to the magnetoresistance of the
sample and detection coil.\cite{Altarawneh2} However, when the temperature
is below $T_{c}$, there is a sudden increase of $\Delta F(H)$ which deviates
from the background signal. This corresponds to entry to the mixed state.
Moreover, with decreasing temperature, the inflexion points of the $\Delta
F(H)$ curves shift to higher field for both field directions, consistent
with the higher $\mu _{0}H_{c2}(T)$ at lower temperature. The temperature
dependence of $\mu _{0}H_{c2}(T)$ for H$\parallel $(101) and H$\perp $(101)
is determined from the intersections of $\Delta F(H)$ curves between the
extrapolated slopes of the rf signals below inflexion points and the
normal-state backgrounds ($T$ = 10 K) (insets in Fig. 1(a) and (b)).\cite%
{Altarawneh2} The difference between this and other criterion, e.g. the
intersection of extrapolated slopes below and above inflexion points in each
$\Delta F(H)$ curve, is taken as the error bar of $\mu _{0}H_{c2}(T)$.

In order to compare the upper critical fields determined from different
measurement methods, the $\mu _{0}H_{c2}(T)$ obtained from $\Delta F(T,H)$
curves and $\rho (T,H)$ data with different criteria are plotted together
(Fig. 2(a)).\cite{Lei HC} In the low field region, the temperature
dependence of $\mu _{0}H_{c2}(T)$ determined from the rf shift is almost
linear with slight upturn near $T_{c}$($H$ = 0 T). This is close to the $\mu
_{0}H_{c2,zero}(T)$ determined from 10\% $\rho _{n}(T,H)$. It is consistent
with the results reported in the literature.\cite{Mun} Assuming $\mu
_{0}H_{c2}(T$ = 0.35 K$)$ $\approx $ $\mu _{0}H_{c2}(0)$, the zero
temperature limit of upper critical fields are 17.4(2) and 19.7(4) T for H$%
\parallel $(101) and H$\perp $(101), respectively. On the other hand,
according to the Werthamer-Helfand-Hohenberg (WHH) theory, orbital pair
breaking field $\mu _{0}H_{c2}(0)$ = -0.693$T_{c}$($d\mu
_{0}H_{c2}/dT|_{T_{c}}$),\cite{Werthamer}\ and using the initial slopes $%
d\mu _{0}H_{c2}/dT|_{T_{c}}$ at low fields obtained from $\rho (T,H)$ data
(-2.54(4) T/K for H$\parallel $(101) and -2.55(4) T/K for H$\perp $(101))
with $T_{c}$ = 8.7 K,\cite{Lei HC} we obtain the $\mu _{0}H_{c2}(0)$ are
15.3(2) and 15.4(2) T for H$\parallel $(101) and H$\perp $(101),
respectively. This is smaller than experimental results. The deviation from
WHH model is clearly seen in Fig. 2(b), where the $\mu _{0}H_{c2}(T)$
becomes gradually larger than expected values from theory. The enhancement
of the $\mu _{0}H_{c2}$ in the low temperature and high field region implies
that multiband effect are not negligible. On the other hand, assuming the
electron-phonon coupling parameter $\lambda _{e-ph}$ = 0.5 (typical value
for weak-coupling BCS superconductors),\cite{Allen} the Pauli limiting field
$\mu _{0}H_{p}(0)$ = 1.86$T_{c}(1+\lambda _{e-ph})^{1/2}$ is 19.8 T.\cite%
{Orlando} This is nearly the same as the $\mu _{0}H_{c2,H\perp (101)}$($T=$
0.35 K) and larger than values for H$\parallel $(101) or the orbital pair
breaking fields. It suggests that the spin-paramagnetic effect might also
have some influence on the upper critical fields. This is rather different
from other FeCh-11 superconductors where the Pauli limiting fields are much
smaller than orbital pair-breaking fields and therefore the
spin-paramagnetic effect governs $\mu _{0}H_{c2}(T)$.\cite{Lei HC1}$^{,}$%
\cite{Lei HC2}

\begin{figure}[tbp]
\centerline{\includegraphics[scale=0.48]{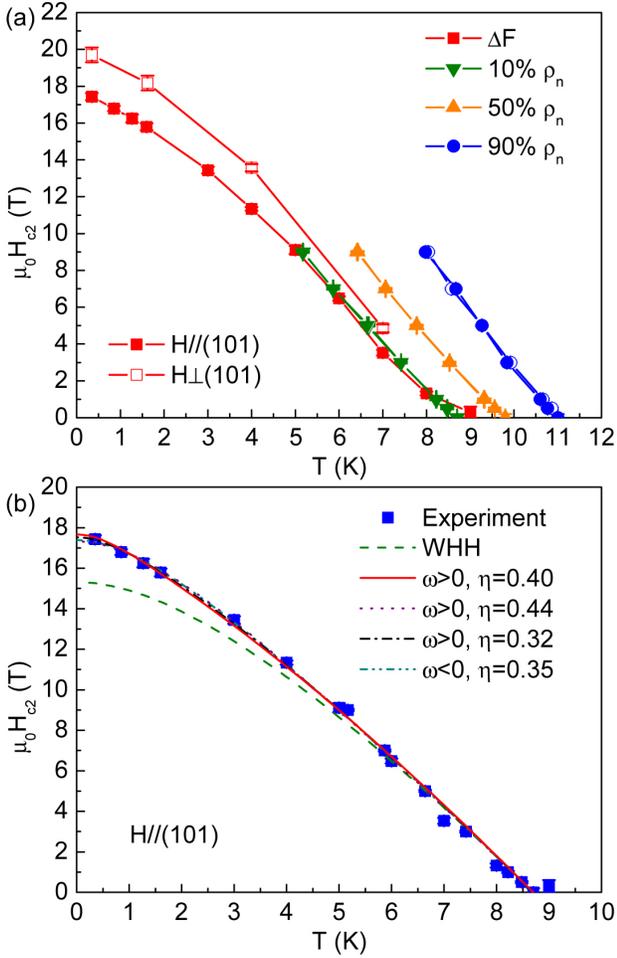}} \vspace*{-0.3cm}
\caption{(a) Temperature dependence of $\protect\mu _{0}H_{c2}(T)$ of $%
\protect\beta $-FeSe single crystal for H$\parallel $(101) (closed symbols)
and H$\perp $(101) (open symbols) obtained from $\protect\rho (T)$ and $%
\Delta F$ curves. (b) Fits of $\protect\mu _{0}H_{c2}(T)$ for H$\parallel $%
(101) using eq. (1) for different pairing scenarios: (1) WHH; (2) $\protect%
\varpi $ $>$ 0, $\protect\lambda _{11}$ = 0.241, $\protect\lambda _{22}$ =
0.195, $\protect\lambda _{12}$ = $\protect\lambda _{21}$ = 0.01, $\protect%
\eta $ = D$_{2}$/D$_{1}$ = 0.40; (3) $\protect\varpi $ $>$ 0, $\protect%
\lambda _{11}$ = $\protect\lambda _{22}$ = 0.5, $\protect\lambda _{12}$ = $%
\protect\lambda _{21}$ = 0.25, $\protect\eta $ = D$_{2}$/D$_{1}$ = 0.44; (4)
$\protect\varpi $ $>$ 0, $\protect\lambda _{11}$ = 0.8 $\protect\lambda %
_{22} $ = 0.34, $\protect\lambda _{12}$ = $\protect\lambda _{21}$ = 0.18, $%
\protect\eta $ = D$_{2}$/D$_{1}$ = 0.32; and (5) $\protect\varpi $ $<$ 0, $%
\protect\lambda _{11}$ = $\protect\lambda _{22}$ = 0.49, $\protect\lambda %
_{12}$ = $\protect\lambda _{21}$ = 0.5, $\protect\eta $ = D$_{2}$/D$_{1}$ =
0.35;}
\end{figure}

According to the two-band BCS model in the dirty limit with orbital pair
breaking and negligible interband scattering,\cite{Gurevich} $\mu _{0}H_{c2}$
is given by

\begin{multline}
a_{0}[\text{ln}t+U(h)][\text{ln}t+U(\eta h)]+a_{2}[\text{ln}t+U(\eta h)] \\
+a_{1}[\text{ln}t+U(h)]=0
\end{multline}

where $t=T/T_{c}$, $U(x)=\psi (1/2+x)-\psi (x)$, $\psi (x)$ the digamma
function, $\eta =D_{2}/D_{1}$, $D_{1}$ and $D_{2}$ are intraband
diffusivities of the bands 1 and 2, $h=H_{c2}D_{1}/(2\phi _{0}T)$, $\phi _{0}
$ the magnetic flux quantum. $a_{0}$, $a_{1}$, and $a_{2}$ are constants
described with intraband- and interband coupling strength, $a_{0}=2\varpi
/\lambda _{0}$, $a_{1}=1+\lambda _{-}/\lambda _{0}$, and $a_{1}=1-\lambda
_{-}/\lambda _{0}$, where $\varpi =\lambda _{11}\lambda _{22}-\lambda
_{12}\lambda _{21}$, $\lambda _{0}=(\lambda _{-}^{2}+4\lambda _{12}\lambda
_{21})^{1/2}$, and $\lambda _{-}=\lambda _{11}-\lambda _{22}$. Terms $%
\lambda _{11}$ and $\lambda _{22}$ are the intraband couplings in the bands
1 and 2 and $\lambda _{12}$ and $\lambda _{21}$ describe the interband
couplings between bands 1 and 2. It should be noted that if $\eta $ = 1, eq.
(1) will reduce to the simplified WHH equation for single-band dirty
superconductors.\cite{Werthamer} By using the coupling constants determined
from an $\mu SR$ experiment with very small interband coupling,\cite%
{Khasanov} the combined $\mu _{0}H_{c2,H\parallel (101)}(T)$ data from both
rf and resistivity measurements can be very well explained (Fig. 2(a) fit
lines). The ratio of band diffusivities is $\eta $ = 0.40, which is similar
to the value of FeAs-122 but much larger than that of other two-band
iron-based superconductors, such as FeAs-1111.\cite{Hunte}$^{-}$\cite{Baily}
With current coupling constants, it leads to the similar shape of $\mu
_{0}H_{c2,H\parallel (101)}(T)$ when compared to the FeAs-122,\cite{Baily}
but significantly different from FeAs-1111 where there is an obvious upturn
at low temperature.\cite{Hunte}$^{,}$\cite{Jaroszynski} We have also
performed fits for different values of coupling constants:\cite{Hunte}$^{-}$%
\cite{Baily} (1) dominant intraband coupling, $\varpi >0$ and (2) dominant
interband coupling, $\varpi <0$. The different sets of fitting parameters
result in almost identical result, fitting the experimental data well (Fig.
2(b)). The derived $\eta $ is in the range of 0.32-0.44, suggesting that the
fitting results are insensitive to the choice of coupling constants. Thus,
either interband and intraband coupling strength are comparable or their
difference is below the resolution of our experiment.

\begin{figure}[tbp]
\centerline{\includegraphics[scale=0.48]{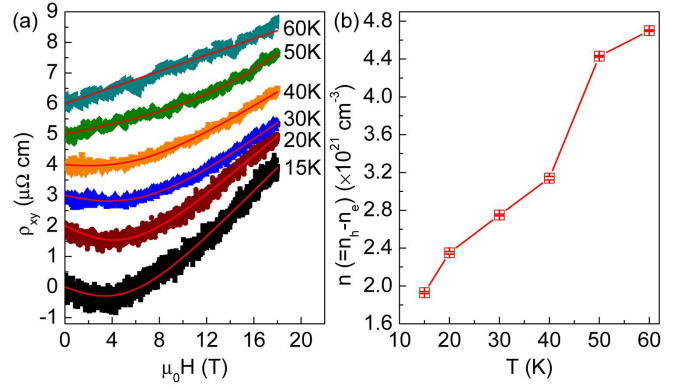}} \vspace*{-0.3cm}
\caption{(a) Field dependence of $\protect\rho _{xy}(H)$ at various
temperatures. Solid lines are the fitting results using eq. (2). for $T<$ 60
K and single-band model for $T=$ 60 K. In order to exhibit data clearly, the
$\protect\rho _{xy}(H)$ at different temperatures are shifted along vertical
axis with certain values. (b) Temperature dependence of carrier density $%
n(=n_{h}-n_{e})$ of $\protect\beta $-FeSe crystal. }
\end{figure}

In order to further investigate multiband characteristics in $\beta $-FeSe,
we studied the Hall effect of $\beta $-FeSe (Fig. 3). According to the band
calculations, at least four bands originated from Fe 3d orbitals cross the
Fermi level.\cite{Subedi}\ Two bands are hole type and the other two are
electron type.\cite{Subedi}\ We use a simplified two-carrier model including
one electron type with electron density $n_{e}$ and mobility $\mu _{e}$ and
one hole type with hole density $n_{h}$ and mobility $\mu _{h}$. According
to the classical expression for the Hall coefficient including both electron
and hole type carriers,\cite{Smith}

\begin{multline}
\rho _{xy}/\mu _{0}H= \\
R_{H}=\frac{1}{e}\frac{(\mu _{h}^{2}n_{h}-\mu _{e}^{2}n_{e})+(\mu _{h}\mu
_{e})^{2}(\mu _{0}H)^{2}(n_{h}-n_{e})}{(\mu _{e}n_{h}+\mu
_{h}n_{e})^{2}+(\mu _{h}\mu _{e})^{2}(\mu _{0}H)^{2}(n_{h}-n_{e})^{2}}
\end{multline}

Once there are two carrier types present, the field dependence of $\rho
_{xy}(H)$ will become nonlinear. Moreover, eq. (2) gives $R_{H}=e^{-1}\cdot
(\mu _{h}^{2}n_{h}-\mu _{e}^{2}n_{e})/(\mu _{e}n_{h}+\mu _{h}n_{e})^{2}$
when $\mu _{0}H\rightarrow $ 0, and $R_{H}=e^{-1}\cdot 1/(n_{h}-n_{e})$ when
$\mu _{0}H\rightarrow $ $\infty $. As shown in inset (a) of Fig. 3, $\rho
_{xy}(H)$ is positive and almost linear in $\mu _{0}H$ at T = 60 K,
indicating the hole type carrier is dominant. However, $\rho _{xy}(H)$
exhibits obvious nonlinear behavior below 50 K and even changes sign in low
fields at 15 K (inset (b) in Fig. 3). This is a signature of coexistence of
electron and hole type carriers. The $\rho _{xy}(H)$ can be described very
well using a linear relation for $T$ $=$ 60 K and eq. (2) for $T$ $\leqslant
$ 50 K as shown with the solid fit lines in the inset (a) and (b) of Fig. 3.
The obtained carrier density $n(=n_{h}-n_{e})$ changes from 1.93$\times $10$%
^{21}$ cm$^{-3}$ (15 K) to 4.7$\times $10$^{21}$ cm$^{-3}$ (60 K) gradually.
The change of sign of $\rho _{xy}(H)$ in the low field region at 15 K
indicates $(\mu _{h}^{2}n_{h}-\mu _{e}^{2}n_{e})$ $<$ 0. Because $%
n_{h}-n_{e}>0$ at higher field, it indicates that the $\mu _{e}>\mu _{h}$ at
low temperature, consistent with the band structure calculation results.\cite%
{Subedi} Moreover, the negative Seebeck coefficients in $\beta $-FeSe below $%
\sim $ 250 K also confirm that the electron band is dominant at low
temperature.\cite{McQueen}

\begin{figure}[tbp]
\centerline{\includegraphics[scale=0.48]{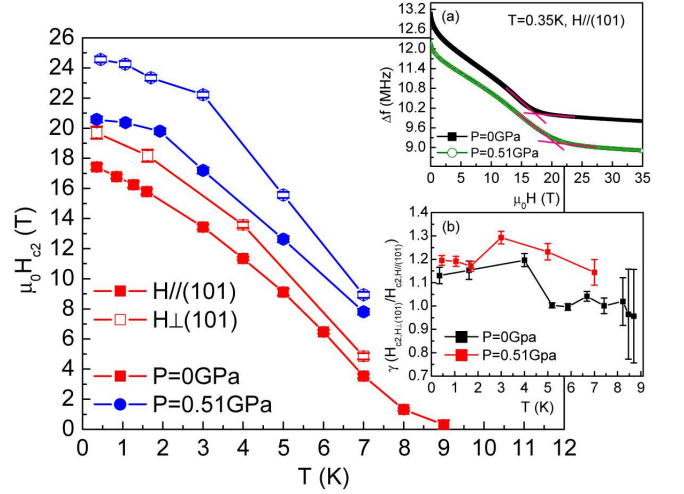}} \vspace*{-0.3cm}
\caption{(a) Temperature dependence of $\protect\mu _{0}H_{c2}(T)$ for H$%
\parallel $(101) (closed symbols) and H$\perp $(101) (open symbols) at
ambient pressure and $P$ = 0.51 GPa obtained from $\Delta F$ curves. Inset
(a) field dependence of $\Delta F$ at 0 and 0.51 GPa for H$\parallel $(101)
at T = 0.35 K. Inset (b) The temperature dependence of the anisotropy of $%
\protect\mu _{0}H_{c2}(P,T)$ at $P$ = 0 and 0.51 Gpa.}
\end{figure}

Since there is remarkable pressure effect on $T_{c}$ for $\beta $-FeSe,\cite%
{Mizuguchi2}$^{,}$\cite{Medvedev} it is instructive to study the pressure
dependence of upper critical fields. As shown in the inset of Fig. 4, under
pressure ($P$ = 0.51 GPa), the inflexion point of $\Delta F(H)$ curve shifts
to higher field when compared to the ambient pressure curve, suggesting that
the $\mu _{0}H_{c2}(T)$ is enhanced with pressure. It is consistent with the
significant positive pressure effect of $T_{c}$ for $\beta $-FeSe.\cite%
{Mizuguchi2}$^{,}$\cite{Medvedev} The temperature dependence of $\mu
_{0}H_{c2}(T)$ for H$\parallel $(101) and H$\perp $(101) shows that the
upper critical fields for both field directions are enhanced in the whole
measured temperature region under pressure. The $\mu _{0}H_{c2}(T=$ 0.45 K$)$
for H$\perp $(101) is about 24.6 T, close to the estimated value at 1.48 GPa
using linear extrapolation.\cite{Mizuguchi2} It suggests that $\mu
_{0}H_{c2}(0)$ at 0.51 GPa should be larger than linear-extrapolated value.
This could originate from the difference in sample purity between our single
crystals and polycrystals or intrinsic multiband effect.\ As shown in the
inset (b) of Fig. 4, at ambient pressure, the anisotropy of $\mu
_{0}H_{c2}(T)$, $\gamma $($P$ = 0 GPa$,T$) = $\mu _{0}H_{c2,H\perp (101)}(P$
= 0 GPa$,T)/\mu _{0}H_{c2,H\parallel (101)}$($P$ = 0 GPa$,T$), is smaller
than in other iron based superconductors, especially at high temperature.
Moreover, the temperature dependence of $\gamma $($P,T$) increases at high
temperature and decreases when $T\ll T_{c}$, which is different from other
iron based superconductors in which the $\gamma $($T$) usually decreases
with temperature. The increase of $\gamma $($P,T$) with temperature has also
been observed in two-band superconductor MgB$_{2}$. This may be due to the
higher contribution of the band with lower band anisotropy at low
temperature. The decrease of $\gamma $($P,T$) with temperature when
temperature is far from Tc may be related to the possible spin-paramagnetic
effect. On the other hand, under pressure, the $\gamma $($P$ = 0.51 GPa$,T$)
increases when compared to the value at ambient pressure. This could
originate from the pressure-induced Fermi surface changes that increase the
anisotropy of Fermi velocity (diffusivity) of dominant band. It should be
noted that in order to study the anisotropy and pressure evolution of $\mu
_{0}H_{c2}(T)$ more clearly, the pressure dependence of $\mu _{0}H_{c2}(T)$
along crystallographic axes should be measured in the future.

\section{Conclusion}

In summary, we studied the upper critical field of $\beta $-FeSe crystals.
The results indicate that the two band effects dominate the $\mu
_{0}H_{c2}(T)$, with possible influence of spin-paramagnetic effect. A
nonlinear field dependence of $\rho _{xy}(H)$ at low temperature also
confirms the existence of multiple bands in electronic transport. The
dominant carriers are hole-type in high field but electron type carriers
become important in low field due to either increased carrier density or
enhanced mobility. The $\mu _{0}H_{c2}(T)$ is enhanced for both field
directions and the anisotropy of $\mu _{0}H_{c2}(0)$ is also increased under
pressure.

\section{Acknowledgements}

Work at Brookhaven is supported by the U.S. DOE under Contract No.
DE-AC02-98CH10886 (R. Hu and H. Ryu) and in part by the Center for Emergent
Superconductivity, an Energy Frontier Research Center funded by the U.S.
DOE, Office for Basic Energy Science (H. Lei and C. Petrovic). Work at the
National High Magnetic Field Laboratory is supported by the DOE NNSA
DEFG52-10NA29659 (S. W.T. and D. G.), by the NSF Cooperative Agreement No.
DMR-0654118 and by the state of Florida.

$^{\ast }$Present address: Center for Nanophysics \& Advanced Materials and
Department of Physics, University of Maryland, College Park MD 20742-4111,
USA.


\begin{thebibliography}{99}
\bibitem{Kamihara} Y. Kamihara, T. Watanabe, M. Hirano, and H. Hosono, J.
Am. Chem. Soc. \textbf{130}, 3296 (2008).

\bibitem{Rotter} M. Rotter, M. Tegel, and D. Johrendt, Phys. Rev. Lett.
\textbf{101}, 107006 (2008).

\bibitem{Wang XC} X. C. Wang, Q. Q. Liu, Y. X. Lv, W. B. Gao, L. X. Yang, R.
C. Yu, F. Y. Li, and C. Q. Jin, Solid State Commun. \textbf{148}, 538 (2008).

\bibitem{Zhu XY} X. Y. Zhu, F. Han, G. Mu, P. Cheng, B. Shen, B. Zeng, and
H.-H. Wen, Phys. Rev. B \textbf{79}, 220512(R) (2009).

\bibitem{Hsu FC} F. C. Hsu, J. Y. Luo, K. W. Yeh, T. K. Chen, T. W. Huang,
P. M. Wu, Y. C. Lee, Y. L. Huang, Y. Y. Chu, D. C. Yan, and M. K. Wu, Proc.
Natl. Acad. Sci. U.S.A. \textbf{105}, 14262 (2008).

\bibitem{Guo} J. Guo, S. Jin, G. Wang, S. Wang, K. Zhu, T. Zhou, M. He, and
X. Chen, Phys. Rev. B \textbf{82}, 180520(R) (2010).

\bibitem{Mazin} I. I. Mazin, D. J. Singh, M. D. Johannes, and M. H. Du,
Phys. Rev. Lett. \textbf{101}, 057003 (2008).

\bibitem{Ding} H. Ding, P. Richard, K. Nakayama, K. Sugawara, T. Arakane, Y.
Sekiba, A. Takayama, S. Souma, T. Sato, T. Takahashi, Z. Wang, X. Dai, Z.
Fang, G. F. Chen, J. L. Luo, and N. L. Wang, EPL \textbf{83}, 47001 (2008).

\bibitem{Hanaguri} T. Hanaguri, S. Niitaka, K. Kuroki, and H. Takagi,
Science \textbf{328}, 474 (2011).

\bibitem{Zhang} Y. Zhang, L. X. Yang, M. Xu, Z. R. Ye, F. Chen, C. He, H. C.
Xu, J. Jiang, B. P. Xie, J. J. Ying, X. F. Wang, X. H. Chen, J. P. Hu, M.
Matsunami, S. Kimura, and D. L. Feng, Nature Mater. \textbf{10}, 273 (2011).

\bibitem{Cruz} C. de la Cruz, Q. Huang, J. W. Lynn, J. Y. Li, W. Ratcliff
II, J. L. Zarestky, H. A. Mook, G. F. Chen, J. L. Luo, N. L. Wang, and P. C.
Dai, Nature \textbf{453}, 899 (2008).

\bibitem{Drew} A. J. Drew, Ch. Niedermayer, P. J. Baker, F. L. Pratt, S. J.
Blundell, T. Lancaster, R. H. Liu, G.Wu, X. H. Chen, I.Watanabe, V. K.
Malik, A. Dubroka, M. R\"{o}ssle, K. W. Kim, C. Baines, and C. Bernhard,
Nature Mater. \textbf{8}, 310 (2009).

\bibitem{Chen H} H. Chen, Y. Ren, Y. Qiu, W. Bao, R. H. Liu, G. Wu, T. Wu,
Y. L. Xie, X. F. Wang, Q. Huang, and X. H. Chen, EPL \textbf{85}, 17006
(2009).

\bibitem{Subedi} A. Subedi, L. Zhang, D. J. Singh, and M. H. Du, Phys. Rev.
B \textbf{78}, 134514 (2008) .

\bibitem{Mizuguchi2} Y. Mizuguchi, F. Tomioka, S. Tsuda, T. Yamaguchi, and
Y. Takano, Appl. Phys. Lett. \textbf{93}, 152505 (2008).

\bibitem{Medvedev} S. Medvedev, T. M. McQueen, I. A. Troyan, T. Palasyuk, M.
I. Eremets, R. J. Cava, S. Naghavi, F. Casper, V. Ksenofontov, G.Wortmann,
and C. Felser, Nature Mater. \textbf{8}, 630 (2008).

\bibitem{Zhang LJ} L. J. Zhang, D. J. Singh, and M. H. Du, Phys. Rev. B
\textbf{79}, 012506 (2009).

\bibitem{Kida} T. Kida, T. Matsunaga, M. Hagiwara, Y. Mizuguchi, Y. Takano,
and K. Kindo, J. Phys. Soc. Jpn \textbf{78}, 113701 (2009).

\bibitem{Lei HC1} Hechang Lei, Rongwei Hu, E. S. Choi, J. B. Warren, and C.
Petrovic, Phys. Rev. B\ \textbf{81}, 094518 (2010).

\bibitem{Lei HC2} Hechang Lei, Rongwei Hu, E. S. Choi, J. B. Warren, and C.
Petrovic, Phys. Rev. B\ \textbf{81}, 184522 (2010).

\bibitem{Hunte} F. Hunte, J. Jaroszynski, A. Gurevich, D. C. Larbalestier,
R. Jin, A. S. Sefat, M. A. McGuire, B. C. Sales, D. K. Christen, and D.
Mandrus, Nature \textbf{453}, 903 (2008).

\bibitem{Jaroszynski} J. Jaroszynski, F. Hunte, L. Balicas, Y.-J. Jo, I. Rai%
\v{c}evi\'{c}, A. Gurevich, D. C. Larbalestier, F. F. Balakirev, L. Fang, P.
Cheng, Y. Jia, and H. H. Wen, Phys. Rev. B\ \textbf{78}, 174523 (2008).

\bibitem{Baily} S. A. Baily, Y. Kohama, H. Hiramatsu, B. Maiorov, F. F.
Balakirev, M. Hirano, and H. Hosono, Phys. Rev. Lett. \textbf{102}, 117004
(2009).

\bibitem{Khasanov} R. Khasanov, M. Bendele, A. Amato, K. Conder, H. Keller,
H.-H. Klauss, H. Luetkens, and E. Pomjakushina, Phys. Rev. Lett. \textbf{104}%
, 087004 (2010).

\bibitem{Hu RW} Rongwei Hu, Hechang Lei, M. Abeykoon, E. S. Bozin, S. J. L.
Billinge, J. B. Warren, T. Siegrist, and C. Petrovic, Phys. Rev. B. \textbf{%
83}, 224502 (2011).

\bibitem{Mielke} C. Mielke, J. Singleton, M.-S. Nam, N. Harrison, C. C.
Agosta, B. Fravel, and L. K. Montgomery, J. Phys. Condens. Matter \textbf{13}%
, 8325 (2001).

\bibitem{Graf} D. E. Graf, R. L. Stillwell, K. M. Purcell, and S. W. Tozer,
High Pressure Res. \textbf{31}, 533 (2011).

\bibitem{Barnett} J. D. Barnett, S. Block, and G.J. Piermarini, Rev. Sci.
Instr. \textbf{44}, 1 (1973).

\bibitem{Coffey} T. Coffey, Z. Bayindir, J. F. DeCarolis, M. Bennett, G.
Esper, and C. C. Agosta, Rev. Sci. Instrum. \textbf{71}, 4600 (2000).

\bibitem{Altarawneh} M. M. Altarawneh, C. H. Mielke, and J. S. Brooks, Rev.
Sci. Instrum. \textbf{80}, 066104 (2009).

\bibitem{Altarawneh2} M. M. Altarawneh, K. Collar, C. H. Mielke, N. Ni, S.
L. Bud'ko, and P. C. Canfield, Phy. Rev. B \textbf{78}, 220505(R) (2008).

\bibitem{Lei HC} Hechang Lei, Rongwei Hu, and C. Petrovic, Phys. Rev. B
\textbf{84}, 014520 (2011).

\bibitem{Mun} E. D. Mun, M. M. Altarawneh, C. H. Mielke, V. S. Zapf, R. Hu,
S. L. Bud'ko, and P. C. Canfield, Phy. Rev. B \textbf{83}, 100514(R) (2011).

\bibitem{Werthamer} N. R. Werthamer, E. Helfand, and P. C. Hohenberg, Phys.
Rev. \textbf{147}, 295 (1966).

\bibitem{Allen} P. B. Allen, in Handbook of Superconductivity, edited by C.
P. Poole, Jr. (Academic Press, New York, 1999) p. 478.

\bibitem{Orlando} T. P. Orlando, E. J. McNiff, Jr., S. Foner, and M. R.
Beasley, Phys. Rev. B \textbf{19}, 4545 (1979).

\bibitem{Gurevich} A. Gurevich, Phys. Rev. B \textbf{67}, 184515 (2003).

\bibitem{Smith} R. A. Smith, \textit{Semiconductors}, Cambridge Univerisity
Press, Cambridge, England, (1978).

\bibitem{Zhang JL} J. L. Zhang, L. Jiao, Y. Chen, H. Q. Yuan, arxiv:
1201.2548 (2011).

\bibitem{Lyard} L. Lyard, P. Szab\'{o}, T. Klein, J. Marcus, C. Marcenat, K.
H. Kim, B.W. Kang, H. S. Lee, and S. I. Lee, Phy. Rev. Lett. \textbf{92},
057001 (2004).

\bibitem{McQueen} T. M. McQueen, Q. Huang, V. Ksenofontov, C. Felser, Q. Xu,
H. Zandbergen, Y. S. Hor, J. Allred, A. J. Williams, D. Qu, J. Checkelsky,
N. P. Ong, and R. J. Cava, Phys. Rev. B \textbf{79}, 014522 (2009).
\end{thebibliography}
\end{document}